\documentstyle[12pt]{article}

\newcommand{\rmsub}[2]{#1_{\rm #2}}
\newcommand{\mathbf}[1]{{\bf #1}}
\newcommand{\mathrm}[1]{{\rm #1}}

\title{\large{DISTANCE ESTIMATION IN COSMOLOGY}}

            \author{\indent M.A. Hendry\\ \\
            \indent Astronomy Centre\\
            \indent University of Sussex\\
            \indent Brighton, BN1 9QH\\
\\
            \indent J.F.L. Simmons\\ \\
            \indent Department of Physics and Astronomy\\
            \indent University of Glasgow\\
            \indent Glasgow, G12 8QQ\\}

\date{}

\begin{document}

\maketitle
\thispagestyle{empty}
\section*{\normalsize{SUMMARY}}
In this paper we outline the framework of mathematical statistics with which
one may study the properties of galaxy distance estimators. We describe,
within this framework, how one may formulate the problem of distance
estimation as a Bayesian inference problem, and highlight the crucial
question of how one incorporates prior information in this approach. We
contrast the Bayesian approach with the classical `frequentist'
treatment of parameter estimation, and illustrate -- with the simple
example of estimating the distance to a single galaxy in a redshift survey --
how one can obtain a significantly different result in the two cases. We
also examine some examples of a Bayesian treatment of distance estimation --
involving the definition of Malmquist corrections -- which have been
applied in recent literature, and discuss the validity of the assumptions
on which such treatments have been based.

\section{\normalsize{INTRODUCTION}}

Recently, the estimation of galaxy distances has assumed great importance
in cosmology. The analysis of large-scale galaxy redshift surveys, used in
conjunction with redshift-independent galaxy distance estimates, can
place powerful constraints on the values of the cosmological parameters
$H_0$ and $\Omega_0$ (c.f. Hendry, 1992b; Dekel, 1994),
and in principle can allow one to test several of the
hypotheses -- including the form of the initial spectrum of density
perturbations, the role of gravity in the growth of structure and the
clustering properties of dark matter -- on which current theories for
the formation of large scale structure in the universe are largely based.
Various methods have been
developed to reconstruct the
density and three-dimensional peculiar velocity field from galaxy
redshift and redshift-distance surveys (c.f. Dekel et al, 1990, 1993;
Simmons, Newsam \& Hendry, 1995;
Rauzy, Lachieze-Rey \& Henriksen, 1994), based upon the ansatz
that the peculiar velocity field is a {\em potential field\/} -- an idea
first developed in the {\sc{potent}} reconstruction method (Bertschinger
\& Dekel, 1989). At the same time, new statistical methods of analysing
surveys which consist of redshifts alone have been developed (c.f.
Lahav et al, 1994; Fisher et al, 1994; Heavens \& Taylor, 1995) based upon the
description of the large scale density and velocity field in terms of
sets of orthogonal functions. One of the biggest current challenges in
this field is to combine in an optimal fashion the results of these two
different methods of analysis, in order to place stronger constraints on
cosmological models and the values of cosmological parameters -- a subject
which would merit an entire article in itself.
In this article we will focus instead only on those issues which concern the
former group of reconstruction methods -- i.e. where one attempts to
obtain redshift-independent distance estimates to galaxies.
\\

Attempts to map the large scale structure of the universe from
redshift-independent galaxy distance estimates have not been without
controversy. For many years considerable debate has been generated over the
precise nature, or indeed the very existence, of galaxy concentrations such
as the `Great Attractor' in the direction of Hydra and Centaurus, for example
(Lynden-Bell et al, 1988; Dressler \& Faber, 1990;
Mathewson, Ford \& Buchhorn, 1992;
Federspiel, Sandage \& Tammann, 1994). A
significant factor fuelling this controversy has been disagreement not
so much over the astrophysical problems of determining `good' galaxy
distance indicators
(although this has undoubtedly played a part also) but rather disagreement
over the equally fundamental question of what {\em statistical\/} methods
one should adopt to analyse the galaxy data. In this paper we attempt to
clarify and place in the open some of the different statistical approaches
which have been adopted in this field of cosmological research, and to
discuss -- within the framework of mathematical statistics -- the different
underlying philosophies upon which (often implicitly) they are based. Our
discussion should be viewed as a general introduction to the problem,
suitable for a reader previously unfamiliar both with the relevant
astronomical details of measuring galaxy distances and with the basic
theory of probability and statistics upon which the topic is founded.
References to more detailed articles, covering both the astronomical and
statistical aspects of the problem, will be given wherever appropriate.
\\

The measurement of the distance of a galaxy, is an example of an
{\em inference\/} problem: i.e. one cannot measure the distance directly
but must infer it from the measurement of some other physical
characteristic, such as the apparent visual magnitude or angular diameter.
If one knew precisely the {\em absolute\/} magnitude or intrinsic
diameter of the galaxy then one could immediately arrive at an exact
determination of the galaxy distance. In early studies of the large scale
distribution and motion of galaxies (c.f. Rubin et al, 1976; Sandage \&
Tammann, 1975a,b) the approach was simply to assume {\em a priori\/} some
fiducial value for this absolute magnitude or diameter and thus infer
galaxy distances on that basis. In practice, however, not all galaxies have
the same absolute magnitude or diameter and so the inference is
statistical in nature. Shortly after these early studies
significant progress was made with the identification of empirical
relationships between absolute magnitude and diameter and other,
distance-independent but directly measurable, physical quantities such as
velocity dispersion or colour (c.f. Faber \& Jackson, 1976; Tully \&
Fisher, 1977; Visvanathan \& Sandage, 1977).
The Tully-Fisher relation, for example, essentially expresses a power law
relationship between the luminosity and the rotation velocity -- as measured
from e.g. the 21cm neutral hydrogen radio emission -- of spiral galaxies.
Thus one measures the 21cm line width of neutral hydrogen for a given
spiral galaxy, applies the Tully-Fisher relation to infer the
absolute magnitude of the galaxy, and then infers the galaxy distance from
its observed apparent magnitude.
\\

In the past decade the Tully-Fisher, and other similar,
relations have been
further refined and placed upon a firmer theoretical footing,
(Pierce \& Tully, 1988; Salucci, Frenk \& Persic, 1993; Hendry et al, 1995)
but they
still contain a significant degree of intrinsic scatter and so do not
provide an exact determination of absolute magnitude or diameter.
Hence, the galaxy distance inferred from such a relation is still
inherently statistical. In the language of mathematical statistics, the
intrinsic scatter of the relation means that we can construct only an
{\em estimator\/} of the galaxy distance, and that distance estimator will
itself be subject to error. More formally, the distance estimator is a
random variable with a definite distribution function, or equivalently
probability density function (pdf), and {\em a fortiori\/} mean and
variance.
\\

Unfortunately there is no unique way to construct distance estimators.
One can make a choice of distance estimator which has certain desirable
properties, the most obvious being that its distribution should have
a small `spread', or variance; on average over many realisations the
estimator should give the true distance of the galaxy; and the
estimator should use all of the information about the galaxy
distance available in the data. These rather loosely defined properties
have their corresponding rigorous definitions in the statistical
literature, and these are referred to as {\em efficiency\/},
{\em unbiasedness\/} and {\em sufficiency\/} respectively.
\\

One should remark that when measurement errors and
intrinsic variability are small in the physical system which one is
modelling, then the adoption of a broad class of different statistical
methods -- or even different statistical philosophies -- in testing models
from observational data will usually make little difference to one's
conclusions. Large discrepancies in the conclusions reached by various
authors in the literature concerning the estimated distances of galaxies
and clusters therefore arise primarily because of large intrinsic
uncertainties inherent to the data.
In other words, galaxy distance indicators are {\em noisy\/}, with typical
distance errors from, e.g., the Tully-Fisher relation of around $20\%$ or
larger to individual galaxies. It is this fact which makes the question of
how one approaches the problem of choosing the `best' galaxy distance
estimator a non-trivial, and an extremely important, one. The typical size
of distance errors has led many cosmologists to attempt to incorporate
prior information on the distribution of galaxy distances when defining
distance estimators, with the aim of reducing the uncertainty in the
final estimate. All examples of this approach can be traced back to what
is termed in the statistical literature as a {\em Bayesian\/} treatment of
the problem of distance estimation, although references in the cosmology and
astronomy literature have often not explicitly used the term `Bayesian',
nor indeed used wholly orthodox Bayesian methods, in their description
of the problem. There are indeed some difficulties with this approach.
One the one hand there are philosophical and methodological problems
that have long been recognised and debated by statisticians
(c.f. Kendall \& Stuart, 1963; Mood \& Graybill, 1974; von Mises, 1957;
Feigelson \& Babul, 1992) which
go to the root definitions and concepts in the theory of probability.
On the other, there is often no clear-cut way of deciding upon the
nature of prior information one can justifiably use. This paper is not
the appropriate place to discuss either of these questions in any great
depth. We would like to emphasise here, however, the principle employed
in Bayesian inference problems in the general statistics literature:
that results which depend heavily on the choice of prior information
should be treated with caution.
\\

Whilst the problem of galaxy distance estimation raises certain
statistical issues which are somewhat unique to astronomy --
in particular the important role of observational selection effects and the
modelling of the physical processes underlying the various distance
relations which are applied to galaxies -- the fundamental concepts are
{\em precisely\/} the same as one finds in the general statistical
literature on inference problems and estimation. It seems sensible,
therefore, for cosmologists to make full use of the `machinery' --
the definitions, notation and general results -- developed by
statisticians for tackling such problems. In this paper, as in our
earlier papers on this subject, we shall attempt to adhere to this
practice.
\\

The structure of this paper is as follows. In section 2 we discuss
in more detail the nature of distance estimators, placing our
discussion in the rigorous context of mathematical statistics and
introducing the appropriate notation and conventions. We go on to
discuss the role of prior information, to explain the concepts of
a `Bayesian' approach to estimation problems, and to examine the
relationship between Bayesian and more orthodox or `frequentist'
approaches. We show, by means of the simple example of estimating
the distances to galaxies in a single catalogue, how a Bayesian
and frequentist approach will yield different results. In section 3
we discuss the various galaxy distance estimators which have been
used in recent literature, drawing particular attention to the
statistical `philosophy'
(i.e Bayesian or frequentist) upon which
they are based, the validity of the assumptions inherent in their
definition, and the extent to which they can be regarded as `good'
estimators -- in the sense of e.g. unbiasedness, efficiency and
sufficency, as introduced above. Finally we discuss the practical
outcomes of using these different estimators for determining
distances to individual galaxies and clusters and in the analysis of
the peculiar velocity and density field by, e.g., the {\sc{potent}}
based methods mentioned above.

\section{\normalsize{STATISTICAL PROPERTIES OF DISTANCE ESTIMATORS}}

One of the purposes of this section is to clarify our notation and
statistical approach for the benefit of the reader previously
unacquainted with the general statistics literature. In the interests
of brevity we shall present here only the essential ideas and omit
unnecessary detail, perhaps at the risk of appearing simplistic. A
more thorough, and wholly rigorous, treatment of the mathematical
foundations of parameter estimation can be found in a large number
of textbooks on probability and statistics (c.f. Hoel, 1962;
Kendall \& Stuart, 1963; Mood \& Graybill, 1974; Hogg \& Craig, 1978)

\section*{\normalsize{What Is an Estimator?}}

In rough terms, an estimator of some unknown parameter is a rule based
on statistical data -- i.e. a random sample drawn from some underlying
population -- for estimating the value of that parameter. If the
parameter of interest is $q$ then we shall write ${\hat{\bf{q}}}$ to
denote an estimator of $q$, following the standard
statistical convention. Note that ${\hat{\bf{q}}}$ is written
in bold face to indicate the fact that it is a {\em random\/}
or {\em statistical variable\/}
(since it is a function of data which are themselves statistical
variables), again in keeping with the standard practice in the
literature. One cannot, of course, expect ${\hat{\bf{q}}}$ to take
on the true value of $q$, $q_0$ say, for {\em every\/} set of statistical
data, but we would regard an estimator as `good' if it tends to yield the
value $q_0$ `on average', or `in the long run' -- rather vague
statements which can be quantified in terms of the {\em bias\/} and
{\em loss function\/} associated with the estimator chosen, as we
discuss below.
\\

By way of an illustrative example, a simple galaxy distance
estimator could be constructed only from the
obaserved apparent magnitude of a galaxy (c.f. Hendry \& Simmons, 1990;
Hendry, 1992a). Thus we may write
%
\begin{equation}
{\bf{m}} - {\bf{M}} = 5 \log r + 25
\end{equation}
where $r$ is the true distance, measured in Mpc, and ${\bf{m}}$
and ${\bf{M}}$ denote
the apparent and absolute magnitude of the galaxy respectively. Of course
the actual distance of the galaxy can only be obtained if there is no
error on the measured value of ${\bf{m}}$ and if ${\bf{M}}$ is known. We can
estimate $r$, however, by making some assumption about the value of
${\bf{M}}$ (for
simplicity we shall ignore any error on ${\bf{m}}$ in this discussion) and
solving for $r$ in equation (1). Suppose we take the value of ${\bf{M}}$ to
be the mean value of absolute magnitude, $M_0$ say, for the underlying
population of all galaxies of a certain Hubble type. We thus obtain an
estimator of log distance, viz
%
\begin{equation}
{\widehat{\log {\bf{r}}}} = 0.2 ( {\bf{m}} - M_0 - 25)
\end{equation}
Here the hat indicates an estimator. If we consider that the
galaxy has been randomly selected from an imaginary population of
galaxies all at the same distance, but with different absolute
magnitudes, then ${\widehat{\log {\bf{r}}}}$ must be considered to be
a statistical variable, as noted previously. The statistical properties
of ${\widehat{\log {\bf{r}}}}$ depend on the galaxy luminosity function
and on the selection function which determines whether a galaxy will or
will not be observed at true distance, $r$. It follows from equations
(1) and (2) that we may write
%
\begin{equation}
{\widehat{\log {\bf{r}}}} = \log r + 0.2 ( {\bf{M}} - M_0 )
\end{equation}
For brevity we shall in future refer to $\log r$ as $w$,
and ${\widehat{\log {\bf{r}}}}$ as ${\hat{\bf{w}}}$.
\\

In general, the underlying pdf {\em before\/} selection for
${\bf{M}}$ is not known. This pdf is usually assumed to be independent
of position and is just the luminosity function (LF) of ${\bf{M}}$,
written $\Psi({\bf{M}})$. The distribution,
$\Psi_{\rm{obs}}({\bf{M}}|r)$,
of ${\bf{M}}$ for observable galaxies at actual distance, $r$,
will depend upon the
selection functio\rm{n a}nd indeed also on $r$ (although for simplicity
we assume here no dependence on direction). Once this pdf,
$\Psi_{\rm{obs}}({\bf{M}}|r)$, is given the pdf of any function of the
random variable, ${\bf{M}}$, may be determined. In particular the pdf
of ${\hat{\bf{w}}}$ defined by equation (3) may easily be found.
Note that while ${\hat{\bf{w}}}$ itself does not depend on
$w$, the pdf of ${\hat{\bf{w}}}$ {\em does\/} depend upon
the true value of the parameter, as one might expect.

\section*{\normalsize{Biased and Unbiased Estimators}}

The mean, or {\em expected\/}, value of a random variable associated with
a galaxy may be taken with respect to either the observable
or the intrinsic galaxy distribution. We shall almost invariably
consider the expectation with respect to the observable distribution in
this paper. Thus the estimator of log distance, ${\hat{\bf{w}}}$,
is defined to be unbiased if
%
\begin{equation}
E ( {\hat{\bf{w}}} | w ) = w
\end{equation}
where the expectation value of any function, $f({\bf{M}})$, of
${\bf{M}}$ is defined as
%
\begin{equation}
E [ f({\bf{M}}) | w ] = \int \! f({\bf{M}})
\Psi ({\bf{M}} | w) d{\bf{M}}
\end{equation}

The bias, $B(w)$, is defined as
%
\begin{equation}
B(w) = E [ {\hat{\bf{w}}} | w ] - w
\end{equation}

When a galaxy survey is subject to a selection limit on apparent
magnitude, the estimator of log distance given by equation (2)
is {\em biased\/} for all true log distances. Moreover, simply
replacing the mean absolute magnitude, $M_0$, of the underlying
population by some fiducially corrected value, $M_0 + c$, where
$c$ is a constant, cannot eradicate this bias (c.f. Hendry \&
Simmons, 1990). One can apply an iterative procedure --
effectively adding a non-constant correction to $M_0$ -- which
considerably reduces the bias of ${\hat{\bf{w}}}$, although
this procedure does not converge to an unbiased estimator for
all log distance (Hendry, 1992a). It has been shown,
however, (c.f. Schechter, 1980; Hendry \& Simmons, 1994) that
in the case of a relation of Tully-Fisher type -- where one
has an additional observable correlated with absolute magnitude --
if the second observable is free from selection effects then one
{\em can\/} define an estimator which is unbiased at all true
log distances. We return to this issue in section 3.

\section*{\normalsize{Minimum Variance and Efficient Estimators}}

There are obvious advantages in using unbiased estimators: in particular,
for large samples -- e.g. when one is estimating the distance to a
rich cluster of galaxies -- the mean estimated distance for the sample
will also be unbiased, and of course will have decreasing variance as the
sample size increases.
Furthermore, if we are interested in, say, the distribution
of actual distances of a catalogue of galaxies, the
histogram of {\em estimated\/} distances can be readily
deconvolved to yield an estimate of this underlying
distribution of true distances. For biased
estimators this would be more difficult (c.f. Eddington, 1913; Newsam,
Simmons \& Hendry, 1994, 1995). Similarly, in model fitting problems -- the
simplest of which in the present context is e.g. the determination of
the Hubble constant -- we can expect parameter estimation to be much easier
if we begin with unbiased estimators. Unbiasedness is not the only criterion
for choosing an estimator, however. It is also natural to desire the
estimator to have a small variance. The variance, $V(w)$, of an estimator is
defined as
%
\begin{equation}
V(w) = E [ ( {\hat{\bf{w}}} - w )^2 | w ]
\end{equation}

In practice one finds that there is a trade-off between small variance and
small bias, in the sense that if you reduce one then you increase the other.
The Cramer-Rao inequality places a lower bound on the variance for both
biased and unbiased estimators (c.f. Hogg and Craig, 1978; Hendry, 1992a;
Gould, 1995; Zaccheo et al, 1995), and an efficient
estimator is one which {\em attains\/} that lower bound -- i.e. which
is a minimum variance estimator.
\\

In choosing an estimator it is also usually convenient to
introduce a {\em loss function\/}, which essentially quantifies
the `loss', or cost, of making an incorrect estimate of a
parameter. An obvious loss function to consider is
%
\begin{equation}
L({\hat{\bf{w}}},w) = ({\hat{\bf{w}}} - w)^2
\end{equation}

A good estimator should yield low values of the expected loss
for a large range of values of the parameter $w$. This
expected loss is called the {\em risk\/}, i.e.
%
\begin{equation}
R(w) = E [ L({\hat{\bf{w}}},w) ]
\end{equation}

Note that for an unbiased estimator the risk and variance are identical,
but for a biased estimator the risk is always strictly greater than the
variance. Thus, if one has an estimator with small variance but large bias,
this would still result in an estimator of large risk -- indicating that
risk is often the more meaningful quantity in comparing estimators. In
general the bias, variance and risk of an estimator are related by the
following simple expression
%
\begin{equation}
R(w) = V(w) + [ B(w)]^2
\end{equation}

\section*{\normalsize{Sufficiency}}

In estimating the distance of a galaxy one does not generally adopt an
estimator of the simple form of equation (2), which is a function only of
apparent magnitude, but rather makes use of a distance indicator such as the
Tully-Fisher relation which depends upon the the strong correlation between
absolute magnitude and some other distance-independent, directly measurable,
observable. Since the underlying physical relationship in an indicator of this
type is unlikely to depend upon only two variables, one could in principle
construct a distance estimator as a function of an arbitrary number of
observables, or {\em statistics\/}. The bias and risk of such an estimator
would depend, of course, upon how well correlated were the observables. In
Hendry \& Simmons (1994) the general case of estimators formed from three
correlated observables is formulated, and in Kanbur \& Hendry (1995) a
specific example is considered where the addition of a {\em fourth\/}
observable -- the maximum apparent magnitude -- to the period, mean luminosity,
colour relation for Cepheid variable stars does indeed result in a distance
estimator of significantly smaller variance and risk.
\\

A obvious general question to ask, then, is whether there exists a function,
say ${\hat{\bf{w}}}({\bf{x}}_1,...,{\bf{x}}_n)$, of a set of observables,
${\bf{x}}_1,...,{\bf{x}}_n$, which `contains' all of the information about
the true value of $w$. Such a function is known as a {\em sufficient
statistic\/} -- and hence would define a sufficient estimator -- for $w$, and
so should be preferred over another estimator without this property. The
property of sufficiency
can be given a more rigorous mathematical definition in terms of the joint
pdf of ${\bf{x}}_1,...,{\bf{x}}_n$ and ${\hat{\bf{w}}}$ (c.f. Mood \&
Graybill, 1974). Suppose that ${\hat{\bf{w}}}_*({\bf{x}}_1,...,{\bf{x}}_n)$
is another statistic based on the observables,
${\bf{x}}_1,...,{\bf{x}}_n$, which is not a function of
${\hat{\bf{w}}}$. Then ${\hat{\bf{w}}}$ is defined to be sufficient if, for
{\em any\/} such ${\hat{\bf{w}}}_*$, the conditional distribution of
${\hat{\bf{w}}}_*$ given ${\hat{\bf{w}}}$ does {\em not\/} depend on the
true parameter value, $w$.
\\

This definition essentially states that once the value of the sufficient
statistic has been specified, one cannot find any other statistic based on
the same set of observables which gives any further information about the
true value of $w$. In a sense, ${\hat{\bf{w}}}$ `exhausts' all the
information about $w$ that is contained in the observed values of
${\bf{x}}_1,...,{\bf{x}}_n$.

\section*{\normalsize{Bayes' Estimators}}

So far we have said nothing about the incorporation of
prior information in the estimation of galaxy distances. Bayesian
approaches attempt to do precisely this.
\\

A fully fledged Bayesian approach would regard $w$ -- in the above
notation -- not as a parameter, but as a statistical variable. The
probability (more commonly referred to as the {\em likelihood\/}) of this
variable taking any given value would be determined by what is
known as its {\em prior distribution\/}: prior, that is, to the data
that we presently have at hand. In the cosmological setting, therefore,
${\bf{w}}$ -- the log distance of a galaxy -- would be taken to have a prior
distribution before the apparent magnitude or diameter or line width
of this galaxy were measured. This prior distribution would be
based on previous information about the distribution of
galaxies as a whole -- or even preconceptions about this
distribution, such as the assumption that the spatial distribution of
galaxies be uniform. In this case one has to modify the orthodox frequentist
view of probability as a `limit' of relative frequencies and adopt instead a
view of probability as a measure of one's state of knowledge about a random
variable.
\\

The {\em posterior\/} distribution for ${\bf{w}}$, once the data for
a particular galaxy has been taken into account, is then
obtained by applying Bayes' theorem. Suppose one's distance estimator is a
function of two variables, ${\bf{m}}$ and ${\bf{P}}$ -- denoting for example
apparent magnitude and log rotation velocity for the Tully Fisher
relation. Bayes' theorem states that
\begin{equation}
p({\bf{m}},{\bf{P}}|{\bf{w}} ) p({\bf{w}}) =
p({\bf{w}}|{\bf{m}},{\bf{P}})p({\bf{m}},{\bf{P}})
\end{equation}

Taking $p({\bf{m}},{\bf{P}})$ to pe a constant, one obtains the
posterior distribution for ${\bf{w}}$, viz
%
\begin{equation}
p({\bf{w}}|{\bf{m}},{\bf{P}}) = C \, p({\bf{m}},{\bf{P}}|{\bf{w}})
p({\bf{w}})
\end{equation}
where $p({\bf{w}})$ is the prior, $C$ is a normalisation constant and
$p({\bf{m}},{\bf{P}}|{\bf{w}})$ is the conditional probability of
${\bf{m}},{\bf{P}}$ given ${\bf{w}}$.
\\

This approach in itself does not give an estimator of
${\bf{w}}$, which is a statistical variable and
not strictly speaking a parameter in the Bayesian
context, but rather it gives a posterior pdf for ${\bf{w}}$ from which one
may define a {\em Bayes' estimator\/} (c.f. Mood \& Graybill, 1974)
in the following way. A Bayes' estimator, ${\bf{\hat{w}}}_{\rm{bayes}}$
minimises the risk, $R({\bf{w}})$ averaged over the prior
distribution, $p({\bf{w}})$ for ${\bf{w}}$. Thus for a
Bayes estimator the integral
%
\begin{equation}
\int \! R({\bf{w}}) \, p({\bf{w}}) \, d{\bf{w}}
\end{equation}
is a minimum. It can be shown that a Bayes' estimator in fact minimises
the loss function averaged over the distribution for ${\bf{w}}$
conditional on the observed data. Explicitly it minimises
%
\begin{equation}
\int \! L({\bf{\hat{w}}}_{\rm{bayes}},{\bf{w}})
p({\bf{w}}|{\rm{data}})d{\bf{w}}
\end{equation}
from which ${\bf{\hat{w}}}_{\rm{bayes}}$ can be found.
\\

It is instructive to consider a simple example where we are
estimating the log distance, $w$, of a galaxy. Let
us assume that we have already an unbiased (in the sense of equation 6)
`raw' estimator, ${\bf{\hat{w}}}$, based on some distance
indicator, which we shall for expediency take to be
normally distributed about the true log distance ${\bf{w}}$ with variance
$\sigma^2$.
Thus the conditional distribution for ${\bf{\hat{w}}}$ given ${\bf{w}}$ is
%
\begin{equation}
p({\bf{\hat{w}}}|{\bf{w}}) = \frac{1}{\sqrt{2 \pi \sigma}}
\exp [ - \frac{1}{2} (\frac{{\bf{\hat{w}}} - {\bf{w}}}{\sigma})^2 ]
\end{equation}

Let us assume, however, that the galaxy is randomly
selected from some underlying population with true log
distance ${\bf{w}}$ distributed normally about some mean value,
$w_c$, and variance $\sigma^2_c$ -- where the
subscript $c$ refers to the catalogue from which the
galaxy is drawn. This normal distribution is taken to be the prior, so in
the above notation
%
\begin{equation}
p({\bf{w}}) = \frac{1}{\sqrt{2 \pi \sigma_c}}
\exp [ - \frac{1}{2} (\frac{{\bf{w}} - w_c}{\sigma_c})^2 ]
\end{equation}

It is now straightforward to show that the conditional distribution
for ${\bf{w}}$ given the value of ${\bf{\hat{w}}}$ is
normally distributed with mean, ${\bf{w}}_B$ and variance
$\sigma^2_B$ given by
%
\begin{equation}
{\bf{w}}_B = {{{\bf{\hat{w}}} + \beta w_c} \over {1+\beta}}
\end{equation}
and
%
\begin{equation}
{\sigma^2_B} = {{\sigma^2} \over {1+\beta}}
\end{equation}
where $\beta = \sigma^2 / \sigma^2_c$,
from which it follows that
%
\begin{equation}
{\bf{\hat{w}}}_{\rm{bayes}} = {{{\bf{\hat{w}}} + \beta w_c} \over
{1+\beta}}
\end{equation}

The interpretation of this result is very
straightforward. If the variance of the indicator is much
smaller than the population variance of the normal distribution of
true log distance for observable galaxies then $\beta \simeq 0$ and
one obtains essentially the `raw' log distance estimator,
${\bf{\hat{w}}}$, suggested by the indicator. If, on the other
hand, the indicator provides very poor information about
the distance of the galaxy then $\beta$ is very large,
and the Bayes estimator yields approximately $w_c$, the mean true log
distance of the observable galaxies in the catalogue. This simple example
demonstrates that, provided the scatter in one's distance indicator is
sufficiently small, one obtains essentially the same estimator irrespective
of whether one adopts a Bayesian approach or not -- and the estimator is
thus largely insensitive to the prior information. The role of the prior
becomes increasingly important, however -- and the difference between a
Bayesian and frequentist approach becomes more apparent -- as the scatter
in the distance indicator increases.
\\

One can regard equation (19) as defining a {\em correction\/}
to ${\bf{\hat{w}}}$ based on the prior information -- in this case that the
underlying populatoin of true log distance is normally distributed.
Corrections of this type
have come to be known in the cosmology literature as {\em Malmquist\/}
corrections, and in the context of mapping large scale structure they were
initially applied assuming the distribution of galaxies to be spatially
homogeneous (c.f. Lynden-Bell et al, 1988; Dekel et al, 1993) -- just as
the distribution of {\em stars\/} had been assumed homogeneous in the
original analytical treatments of Malmquist (1920, 1922) and
Eddington (1913). Recently, however, attempts have been made to apply more
general, inhomogeneous, Malmquist corrections which address the fact that
the galaxy distribution displays small-scale clustering
(c.f. Landy \& Szalay, 1992; Hudson, 1994; Dekel, 1994; Newsam, Simmons \&
Hendry, 1995; Hudson et al, 1995; Freudling et al, 1995).
We briefly consider some important technical problems
regarding the application of inhomogeneous Malmquist corrections in
section 3. It is worth noting here, however, that an entirely frequentist
approach to distance estimation has the advantage that the definition
of an unbiased estimator is completely independent of the underlying
galaxy true number density -- and hence is unaffected by arguments about
the form of prior distribution which one should adopt.

\section{\normalsize{GALAXY DISTANCE INDICATORS IN RECENT LITERATURE}}

Most redshift-independent methods of estimating galaxy distances which have
featured in the recent cosmological literature are based upon
{\em secondary\/} distance indicators -- which require to be calibrated using
a sample of galaxies in, e.g., a nearby cluster, the distance of which is
already known. Notable exceptions to this have been the recent extension
to beyond the Local Group of the extragalactic distance scale measured from
Cepheid variables and the application of the expanding photosphere method
(EPM) to determine the distances of type II supernovae (SN). Both Cepheids
and type II SN are examples of primary distance indicators which can be
calibrated either locally -- within our own galaxy -- or from theoretical
considerations. For a discussion of the physical basis for these
indicators the reader is referred to, e.g., Kirschner \& Kwan (1974),
Eastman \& Kirschner (1989), Jacoby et al (1992) and references therein. Both
indicators have a small intrinsic dispersion ($\sim 10 - 15\%$ to individual
objects) and are thus considerably less susceptible to the problems which
arise in the definition of Malmquist corrections and sensitivity to the
choice of
prior information (essentially because the $\beta$, the ratio of the estimator
variance to the variance of the underlying population, is small).
This property of course makes both Cepheids and type II SN well suited to the
estimation of the Hubble constant -- either directly or in combination with
other secondary indicators such as type Ia SN (c.f. Saha et al, 1994; 1995).
Indeed, the high estimates of H$_0$ reported in
Freedman et al (1994) and Pierce et al (1994), based on the distance of
Cepheids in Virgo cluster galaxies, and those of Schmidt et al (1992, 1994)
based on the EPM distances of type II SN beyond the Local Supercluster
(and thus less adversely affected by peculiar velocities), provide a
compelling argument in favour of a value of H$_0 \geq
60$kms$^{-1}$Mpc$^{-1}$ -- despite the difficulties of reconciling
these results with astrophysical estimates of the age of the galactic disc
(c.f. van den Bergh 1995; Chamcham \& Hendry, 1995).
\\

Of the secondary distance indicators currently in widespread use, only two
are thought to be sufficiently accurate to make the question of how to best
use prior information essentially unimportant: these are surface brightness
fluctuations (SBF) and the luminosity -- light curve shape relation for
type Ia SN. The former distance indicator, SBF, was pioneered by
Tonry \& Schneider (1988) and is based upon the fact that the
fluctuations -- due to the discreteness of individual
stars -- in surface brightness across the CCD image of a nearby elliptical
galaxy will be larger than those for a more distant galaxy. The physical
basis of SBF and the details of its calibration are described in
Jacoby et al (1992). Relative distances of a typical accuracy of
5\% have been derived to a sample of several hundred ellipticals out to a
redshift of around 6000 kms$^{-1}$ using this indicator (Dressler, 1994).
\\

Type Ia SN have long been recognised as useful `standard candles' since they
are observable to very large distances and have a luminosity function which is
well described a Gaussian distribution of dispersion around 0.5 mag.
(Sandage \& Tammann, 1993; Hamuy et al, 1995). In Vaughan et al (1995), it is
argued that the pre-selection of SN based on a colour criterion reduces this
dispersion to $\sim 0.3$ mag., which -- although a significant improvement --
still represents a typical percentange distance error of around 15\% to an
individual galaxy. In Riess, Press \& Kirschner (1995a,b) however, the
shape of the SN light curve is used to more tightly constrain the peak
luminosity and leads to a typical relative distance error of only 5\% -- small
enough to render Malmqust corrections largely unimportant. This method has been
used both to estimate the Hubble constant and to determine the bulk flow
motion of the Local Group on a scale of $\sim 7000$ kms$^{-1}$, yielding a
motion which is consistent with the COBE measurement of the dipole
anisotropy in the microwave background radiation, but inconsistent with the
dipole motion reported by Lauer \& Postman  (1994), based on the redshifts of
Abell clusters at distance of $8000 - 11000$ kms$^{-1}$.
\\

The vast majority of recent analyses of the peculiar velocity and density
fields, and the estimation of the density parameter $\Omega_0$ using
redshift-independent distance indicators, have
been carried out primarily with the Tully-Fisher (TF) and $D_{\rm{n}}-\sigma$
distance indicator relations for spirals and ellipticals respectively. As we
remarked above, the TF relation essentially expresses a power
law relationship between the luminosity and rotation velocity for spiral
galaxies; the $D_{\rm{n}}-\sigma$ relation similarly expresses a power law
relationship between the central velocity dispersion and isophotal
diameter of early-type galaxies (c.f. Jacoby et al, 1992). Although the
number of galaxy distances estimated by these two relations currently
stands at over 4000 (around a factor of ten larger than the number of
distance estimates from SBF and SN distance indicators), and continues
to grow rapidly each year, both the TF and $D_{\rm{n}}-\sigma$ relations are
considerably more noisy -- with dispersions of around 20\% to individual
galaxies. It is for this reason that the issue of how -- or
indeed if -- one should make use of prior information in the definition of
`optimal' estimators continues to be regarded as of crucial importance when
interpreting the results of applying these distance indicators to analyse
redshift surveys.
\\

Both the TF and $D_{\rm{n}}-\sigma$ relations are usually calibrated by
performing a linear regression on a calibrating sample of
galaxies whose distances are otherwise known. It is instructive to
consider this calibration procedure in more detail, in order
to illustrate some of the statistical pitfalls which may arise, for the
generic example of the TF relation. As before, we denote the log
rotation velocity by ${\bf P}$ and let ${\bf \hat{M}}$ denote the
estimator of absolute magnitude which one derives from the TF relation, from
which one may derive the corresponding `raw' estimator of log distance,
${\hat{\bf{w}}}$, from equation (3) in the obvious way.
Thus, we obtain from the calibration a linear relationship between
${\bf \hat{M}}$ and ${\bf P}$,
\begin{equation}
\label{eq:Mhat}
   {\bf \hat{M}} = \alpha {\bf P} + \beta
\end{equation}
where $\alpha$ and $\beta$ are constants. The choice of {\em which\/} linear
regression is most appropriate is non-trivial when one's survey is subject to
observational selection effects. We can demonstrate this with the following
simple example. Suppose that the intrinsic joint distribution of absolute
magnitude and log(rotation velocity) is a bivariate normal.
Figure 1 shows schematically the scatter in the TF relation in
this case, for a calibrating sample which is free from selection effects --
e.g. a nearby cluster. (More precisely, the ellipse shown is an
isoprobability contour enclosing a given confidence region for ${\bf M}$ and
${\bf P}$). The solid and dotted lines show the linear relationship obtained by
regressing rotation velocities on magnitudes and magnitudes on line widths
respectively.
Thus the dotted line is defined as the expected value of ${\bf M}$ at given
${\bf P}$, while the solid line is defined as the expected value of ${\bf P}$
at given ${\bf M}$. Since in practice one wishes to infer the value of
${\bf M}$ from the measured value of ${\bf P}$, the ${\bf M}$ on ${\bf P}$
regression has been referred to in the literature as defining the `direct'
or `forward' TF relation, while using the ${\bf P}$ on ${\bf M}$ regression
defines the `inverse' TF relation. For the bivariate normal case the
equations of the direct and inverse regression lines are as follows:-
\begin{equation}
\label{eq:MgivP}
   E ({\bf M}|{\bf P}) = \rmsub{M}{0} + \frac{\rho \rmsub{\sigma}{M}}
                        {\rmsub{\sigma}{P}} ({\bf P} - \rmsub{P}{0})
\end{equation}
\begin{equation}
\label{eq:PgivM}
   E ({\bf P}|{\bf M}) = \rmsub{P}{0} + \frac{\rho \rmsub{\sigma}{P}}
                        {\rmsub{\sigma}{M}} ({\bf M} - \rmsub{M}{0})
\end{equation}
where $\rmsub{M}{0}$, $\rmsub{P}{0}$, $\rmsub{\sigma}{M}$, $\rmsub{\sigma}{P}$
and $\rho$ denote the means, dispersions and correlation coefficient of the
bivariate normal distribution of ${\bf M}$ and ${\bf P}$.
Both regression lines can be written in the form of equation (20),
thus defining ${\bf \hat{M}}$ as a function of ${\bf P}$, although of
course the constants $\alpha$ and $\beta$ are different in each case.
Moreover the definition of ${\bf \hat{M}}$ is subtly different in each case.
For the direct regression ${\bf \hat{M}}$ is identified as the mean
absolute magnitude at the observed log line width. For the inverse regression
on the other hand ${\bf \hat{M}}$ is defined such that the observed log line
width is equal to its expected value when ${\bf M} = {\bf \hat{M}}$.
Consequently, as is apparent from their slopes, the direct and inverse
regression lines give rise to markedly different distance
estimators, although it is straightforward to show that in the absence of
selection effects both estimators are unbiased, in the sense defined in
equation (4), above.
\\

\begin{figure}
\vspace{7cm}
\caption{Schematic `Direct' and `Inverse' Tully-Fisher relations for the
case of a nearby, completely sampled, cluster.}
\end{figure}

The situation is very different when we include the effects of observational
selection, however. This is illustrated in Figure 2, which
shows the scatter in the TF relation in a calibrating sample subject to a
sharp cut-off in absolute magnitude -- as would be the case in e.g. a
distant cluster observed in an apparent magnitude-limited survey. We can
see that in this case the slope of the direct regression
of ${\bf M}$ on ${\bf P}$ is substantially changed from that in the nearby
cluster -- indeed the direct regression is no longer linear at all. This means
that if one calibrates the TF relation in the nearby cluster using the direct
regression and then applies this relation to the more distant cluster, one
will systematically underestimate its distance, since the expected value of
${\bf M}$ given ${\bf P}$ in the distant cluster is systematically brighter
than that in the nearby cluster as fainter galaxies progressively `fade out'
due to the magnitude limit. The corresponding `direct', or `M on P', log
distance estimator will therefore be negatively biased.
\\

\begin{figure}
\vspace{7cm}
\caption{Schematic `Direct' and `Inverse' Tully-Fisher relations for
the case of a distant cluster subject to a sharp selection limit on
absolute magnitude.}
\end{figure}

In an important paper Schechter (1980) observed that the slope of the
inverse regression line is unchanged,
irrespective of the completeness
of one's sample, provided that the selection effects are in magnitude only.
We can see that this observation is valid in the simple case considered in
Figure 2. In other words the expected value of ${\bf P}$ given ${\bf M}$ is
unaffected by the selection effects and, therefore, defines an unbiased log
distance estimator. In Hendry \& Simmons (1994), Schechter's result is derived
within the rigorous framework of mathematical statistics, and the
assumptions upon which it is based are generalised. In particular it is
shown that the inverse TF log distance estimator is gaussian and unbiased at
all true log distances provided only that the conditional distribution of
${\bf P}$ given ${\bf M}$ is Gaussian, that $E({\bf P}|{\bf M})$ is a linear
function of ${\bf M}$, and that the sample is not subject to selection on
rotation velocity. Moreover, since the inverse log distance estimator is
Gaussian it will also automatically be a {\em sufficient\/} and
{\em efficient\/} estimator, as defined in section 2. In Hendry (1992a)
it was also shown that when there is no selection on rotation velocity
then the inverse log distance estimator is the {\em only\/} unbiased
estimator which is a linear function of log rotation velocity and apparent
magnitude. In particular, the `orthogonal'
(c.f. Giraud 1987), `bisector' (c.f. Pierce \& Tully 1988) and
`mean' (c.f. Mould et al 1993) regression lines also give rise to estimators
which are biased at all true log distances in this case. A similar
conclusion was also reached in Triay, Lachieze-Rey \& Rauzy (1994).
\\

The unbiased properties of the inverse TF relation have led to its use
in defining a `raw' distance estimator in
a number of different recent analyses of the peculiar velocity field,
including Newsam, Simmons \& Hendry (1995), Freudling et al (1995),
Nusser \& Dekel (1994), Shaya, Tully \& Pierce (1992) and Shaya, Tully \&
Peebles (1995). Its acceptance has been far from universal, however. Part
of the reason for this is that, of course, in practice it is {\em not\/}
the case that galaxy samples are free from selection effects on rotation
velocity. In fact, it is commonly the case that redshift surveys are first
selected on the basis of either apparent diameter or B-band apparent
magnitude, or both, while the TF photometry is then carried out in the
near infra-red, or I-band. This leads to a considerably more complex
selection function, as modelled in Sodre \& Lahav (1993), which in general
renders {\em all\/} linear regressions biased. Essentially this problem
arises because diameter, I-band and B-band magnitude and rotation velocity
are mutually correlated variables, so that the selection on B-band
magnitude and angular diameter `pollutes' the joint distribution of
I-band magnitude and rotation velocity in the TF relation -- thus
effectively rendering the assumptions inherent in deriving the unbiasedness
of the inverse TF relation no longer valid (c.f. Hendry \& Simmons 1994;
Willick 1994).
\\

One {\em can\/} determine the correct slope and zero point of the `direct' TF
relation from a cluster subject to observational selection effects by the
application of straightforward iterative procedure -- thus solving what has
been termed in the literature as the `calibration problem' (c.f. Willick 1994;
Hendry et al, 1995). It is important to recognise, however, that the
corresponding `raw' log distance estimator will {\em still\/} be biased,
in the sense of equation (4), at all true distances if applied to a galaxy
survey subject to magnitude selection effects. This is because the joint
distribution of absolute magnitude and log rotation velocity for
observable galaxies will {\em not\/} be equal to the intrinsic joint
distribution.
\\

Why has the use of the `direct' TF relation in recent literature continued
to be widespread? To understand the reason for this we must first note that
most recent analyses of galaxy distances and peculiar velocities have been
carried out within a Bayesian framework, thus involving the application
of Malmquist corrections to the `raw' log distance estimator. The
motivation for adopting a Bayesian approach (even if the Bayesian nature
of the problem has not always been explicitly acknowledged by authors!)
comes about from the way in which galaxy distance estimates and redshifts
have been combined in the majority of analyses. In both the early `toy'
parametric velocity field models of e.g. Lynden-Bell et al (1988),
Dressler \& Faber (1990), and the more sophisticated
reconstruction methods such as POTENT (c.f. Dekel et al 1990), essentially
galaxies are binned and grouped together and assigned radial peculiar
velocity estimates on the basis of their {\em estimated\/} distance.
The galaxy's actual distance could be radically different, and will depend on
the true spatial distribution of galaxies and the exact nature of the
survey selection function. Clearly galaxies which have small estimated
distance are more likely to have been scattered down from larger true
distances, since a volume element of fixed solid angle increases in size
with true distance; close to the limit of the survey volume, however, this
might no longer be the case, as galaxies scattered from larger true distances
might be too faint to be included in the redshift survey. By requiring that on
average the actual radial coordinate of the galaxy be equal to its estimated
distance, one would also ensure that on average the correct
peculiar velocity would be ascribed to that galaxy's apparent position.
The estimator which satisfies this condition can be defined following the
Bayesian approach outlined for the simple illustrative example of section 2,
and it is straightforward to show that such a `Malmquist corrected'
distance estimator, ${\bf{\hat{r}}}_{\rm{bayes}}$, satisfies the equation
\begin{equation}
{\bf{\hat{r}}}_{\rm{bayes}} = C \, \int \! {\rm{dexp}}
({\bf{\hat{w}}}) p({\bf{\hat{w}}}|{\bf{w}}) \, p({\bf{w}}) \, d{\bf{w}}
\end{equation}
where $C$ is a normalisation constant.
\\

The key point about equation (23) is that -- as before -- the
Bayesian distance estimator
depends upon the prior distribution of true log distance, $p({\bf{w}})$.
There has been no consensus in the literature on which prior one should
adopt. As we mentioned in section 2, in Lynden-Bell et al (1988) the prior
is assumed to correspond to a homogeneous distribution of galaxies --
thus defining homogeneous Malmquist corrections which are a function only
of distance. In Landy \& Szalay (1992), on the other hand, a more general
correction is derived by first estimating $p({\bf{w}})$ from
a spline fit to the histogram of log distance
{\em estimates\/} for the galaxies in the survey, thus in principle taking
into account inhomogeneities in the galaxy distribution. Due to the
sparseness of surveys, however, it is usually necessary to
average the distribution of galaxies over large solid angles, if not all, of
the sky. Therefore, the effects of clustering may still go largely
unaccounted for in the Landy \& Szalay prescription (c.f. Newsam et al, 1994).
In other recent analyses (c.f. Hudson 1994; Hudson et al 1995; Dekel 1994;
Willick 1994; Freudling et al 1995) a different method is proposed for
obtaining the prior distribution -- by
reconstructing the density field of optical or IRAS-selected galaxies
based on redshifts alone, assuming linear or mildly non-linear theory to
adequately describe the gravitational collapse of structure -- smoothed on a
scale of the order of 10 Mpc.
\\

In {\em all\/} of the above analyses the Malmquist corrections are derived
assuming that the conditional distribution of the `raw' log distance
estimator, $p({\bf{\hat{w}}}|{\bf{w}})$, is normally distributed at all
true log distances. As shown in Hendry \& Simmons (1994), this assumption
is invalid when the `raw' estimator is derived from the `Direct' TF
relation. Thus, the formula of Landy \& Szalay will result in an
{\em incorrect\/} Malmquist correction due to the bias of the `Direct'
TF log distance estimator. In general, if the prior distribution of true
log distance is inferred from the
observed distribution of log distance {\em estimates\/}, then one must
apply the formula of Landy \& Szalay using the `Inverse' TF estimator --
which we have seen {\em is\/} normally distributed and unbiased at all
true log distances, subject to the conditions specified above and in
Hendry \& Simmons (1994). A similar conclusion is reached in
Teerikorpi (1993), Feast (1994) and Freudling et al (1995).
\\

It is further shown in Hendry \& Simmons (1994) that the use of the `Direct'
TF relation as the raw log distance estimator in defining general
Malmquist corrections can only be justified if the prior distribution in
equation (23) corresponds to the {\em intrinsic\/} distribution of true log
distance. As a special case of this result, note that the homogeneous
Malmquist correction of Lynden-Bell et al (1988) applied to the
`Direct' TF estimator will therefore be valid provided that the intrinsic
distribution of galaxies is homogeneous. In a similar way, the
inhomogeneous corrections derived in Hudson et al (1995), Dekel (1994) and
Freudling et al (1995), will be valid provided the density field
reconstructed from optical or IRAS-selected surveys corresponds to the
intrinsic distribution of true log distance for the TF galaxies -- in other
words that the selection function of the redshift survey has been adequately
corrected for, and the redshift survey faithfully traces the {\em same\/}
underlying population as the galaxies to which the TF relation is being
applied.
\\

\section{\normalsize{CONCLUSIONS}}

In this paper we have set out to describe -- within the framework of
mathematical statistics -- some of the properties of `optimal' galaxy distance
estimators, including unbiasedness, sufficiency and efficiency. We have
shown that the intrinsic scatter of indicators
such as the Tully-Fisher and $D_{\rm{n}}-\sigma$ relations is sufficiently
large that the question of which statistical philosophy one should adopt
in the analysis of redshift surveys is far from trivial.
In particular we have seen that one may formulate the problem of galaxy
distance estimation as a Bayesian inference problem -- essentially the
approach which has been adopted implicitly in the literature in
defining Malmquist-corrected distance estimators -- but that there is no
general agreement over the issue of how one should then
best make use of prior information on the distribution of true galaxy
distances. In particular, a failure to adequately understand the
properties of the `raw' galaxy distance estimator used can lead to the
definition of invalid Malmquist corrections, as was the case in e.g.
Landy \& Szalay (1992). In Newsam, Simmons \& Hendry
(1995) we show that the use of such invalid corrections can frequently be
worse than applying no corrections at all. A similar conclusion was
reported in Freudling et al (1995), where it was shown that a number of
biases may have gone unresolved in earlier attempts to incorporate prior
information in the definition of distance estimators.
\\

In reality, the issue of defining an `optimal' galaxy distance estimator
is only the first part of the story. In applying the POTENT procedure, for
example, whether or not a distance estimator is biased is not the
crucial question; what is important is to construct an unbiased smoothed
peculiar velocity field. Although there appears some justification as to
why this procedure requires the application of an essentially Bayesian
approach, the Malmquist corrections which this approach entails
are strictly only valid if galaxies are not too sparse, the
gradient of the velocity field is not too large, and the effective radius
of the window function used to smooth the data is not too wide. In
Newsam, Simmons \& Hendry (1995) a Monte-Carlo procedure, involving
the generation of large numbers of `mock' redshift surveys, is devised and
implemented with the purpose of eliminating {\em all\/} biases from the
POTENT-recovered velocity and density fields -- not only those associated
with the scatter of the distance indicators. A similar algorithm may be
adopted for other reconstruction methods, and has the distinct advantage
of being easily adapted to more general (and more realistic!) selection
functions and distance indicators -- involving, e.g., correlations between
three or more observables where a wholly analytic treatment can often be
intractable (c.f. Hendry \& Simmons, 1994). A very similar Monte-Carlo
approach has been adopted in
Freudling et al (1995). These papers serve as an important reminder
that the question of galaxy distance estimation cannot be regarded in
isolation: ultimately the choice of which distance estimator is `optimal'
depends on the context in which the distance estimator is being used.
\\

It is perhaps worthwhile to end on a positive note. The use of redshift
independent galaxy distance indicators in conjunction with redshift surveys
has opened up an exciting -- and highly productive -- `industry' in
cosmology during the past decade or so. Although the statistical problems
arising from the large intrinsic scatter of these indicators are considerable,
the mathematical machinery briefly sketched in this paper equips us with the
necessary tools to address important issues such as their sensitivity to the
choice of prior information. Moreover, the significant recent advances made in
developing and applying more accurate distance indicators, such as surface
brightness fluctuations and the supernova light curve shape method, offer
some further cause for optimism: perhaps within the next decade we will be
able to map the large scale structure of the local universe with sufficient
accuracy that the question of whether one should adopt a Bayesian approach
to the analysis -- and how in detail it should be implemented -- will no
longer be important.
\\

Putting this another way, in the general statistics literature on
Bayesian inference, when one's results are sensitive to the choice of prior
information, one is usually advised to go out in search of better data.
Fortunately for those cosmologists measuring galaxy distances, it appears
that such data are indeed on their way!

\section*{\normalsize{REFERENCES}}
\footnotesize
\begin{itemize}
%
\item{Bertschinger E., Dekel A., 1989, Ap.J. (Lett), {\bf{336}}, L5}
%
\item{Chamcham K., Hendry M.A., 1995, MNRAS, submitted}
%
\item{Dekel A., 1994, Ann. Rev. Astr. Astrophys., 371 }
\item{Dekel A., Bertschinger E., Faber S.M., 1990, Ap.J.,
{\bf{364}}, 349}
\item{Dekel A., Bertschinger E., Yahil A., Strauss M.S.,
Davis M., Huchra J., 1993, Ap.J., {\bf{412}}, 1}
\item{Dressler A., 1994, in
`Cosmic velocity fields', eds. Lachieze-Rey M., Bouchet F., (Editions
Frontieres), 9}
\item{Dressler A., Faber S., 1990, Ap.J., {\bf{354}}, 13}
%
\item{Eastman R.G., Kirschner R.P, 1989, Ap.J., {\bf{347}}, 771}
\item{Eddington A., 1913, MNRAS, {\bf{73}}, 359}
%
\item{Faber S.M., Jackson R.E., 1976, Ap.J., {\bf{204}}, 668}
\item{Feast M., 1994, MNRAS, {\bf{266}}, 255}
\item{Federspiel M., Sandage A., Tammann G.A., 1994, Ap.J., {\bf{430}}, 29}
\item{Fiegelson E., Babul C.J., 1992, `Statistical Challenges in Modern
Astronomy', (Springer-Verlag)}
\item{Fisher K., Davis M., Strauss M., Yahil A., Huchra J.P.,
1994, MNRAS, {\bf{267}}, 927}
\item{Freedman W.L., et al, 1994, Nature, {\bf{371}}, 757}
\item{Freudling W., Da Costa L., Wegner G., Giovanelli R., Haynes M.,
Salzer J., 1995, A.J., in press}
%
\item{Giraud E., 1987, Astron. Astrophys., {\bf{174}}, 23}
\item{Gould A., 1995, Ap.J., {\bf{440}}, 510}
%
\item{Hamuy M., Phillips M.M., Maza J., Suntzeff N., Schommer R.,
Aviles R., 1995, A.J., {\bf{109}}, 1}
\item{Heavens, A.F., Taylor, A.N., 1995, MNRAS, in press}
\item{Hendry M.A., 1992a, PhD Thesis, Univ. of Glasgow}
\item{Hendry M.A., 1992b, Vistas in Astronomy, {\bf{35}}, 239}
\item{Hendry M.A., Rauzy S., Salucci P., Persic M., 1995,
Astrophys. Lett. \& Comm., in press}
\item{Hendry M.A., Simmons J.F.L., 1990, Astron. Astrophys.,
{\bf{237}}, 275}
\item{Hendry M.A., Simmons J.F.L., 1994, Ap.J., {\bf{435}}, 515}
\item{Hoel P., 1962, `An introduction to mathematical statistics',
(Wiley, NY)}
\item{Hogg R., Craig A., 1978, `An introduction to mathematical
statistics', (MacMillan)}
\item{Hudson M.J., 1994, MNRAS, {\bf{266}}, 468}
\item{Hudson M.J., Dekel A., Courteau S., Faber S.M., Willick J.A.,
1995, MNRAS, in press}
%
%
\item{Jacoby G.H., et al, 1992, PASP, {\bf{104}}, 599}
%
\item{Kanbur S.M., Hendry M.A., 1995, Astron. Astrophys., in press}
\item{Kendall M., Stuart A., 1963, `The advanced theory of statistics',
(Haffner, NY)}
\item{Kirschner R.P., Kwan J., 1974, Ap.J., {\bf{193}}, 27}
%
\item{Lahav O., Fisher K.B., Hoffman Y., Scharf C., Zaroubi S.,
1994, Ap.J. (Lett), {\bf{423}}, L93}
\item{Landy S., Szalay A., 1992, Ap.J., {\bf{391}}, 494}
\item{Lauer T.R., Postman M., 1994, Ap.J., {\bf{425}}, 418}
\item{Lynden-Bell D., Faber S.M., Burstein D., Davies R.D., Dressler A.,
Terlevich R.J., Wegner G., 1988, Ap.J., {\bf{326}}, 19}
%
\item{Malmquist K.G., 1920, `Medd. Lund. Astron. Obs.', 20}
\item{Malmquist K.G., 1922, `Medd. Lund.' Ser. I., 100}
\item{Mathewson D.S., Ford V.L., Buchhorn M., 1992, Ap.J. (Lett),
{\bf{389}}, L5}
\item{von Mises R., 1957, `Probability, statistics \& truth', (Allen \&
Unwin)}
\item{Mood A.M., Graybill A.F., 1974, `Introduction to the theory of
statistics', (McGraw-Hill, NY)}
\item{Mould J., Akeson R.L., Bothun G.D., Han M., Huchra J.P., Roth J.,
Schommer R.A., 1993, Ap.J., {\bf{409}}, 14}
%
\item{Newsam A.M., Simmons J.F.L., Hendry M.A., 1994, in
`Cosmic velocity fields', eds. Lachieze-Rey M., Bouchet F., (Editions
Frontieres), 49}
\item{Newsam A.M., Simmons J.F.L., Hendry M.A., 1995, Astron.
Astrophys., {\bf{294}}, 627}
\item{Nusser A., Davis M., 1994, Ap.J. Lett., {\bf{421}}, L1}
%
%
\item{Pierce M.J., Tully R.B., 1988, Ap.J., {\bf{330}}, 579}
\item{Pierce M.J., et al, 1994, Nature, {\bf{371}}, 385}
%
%
\item{Rauzy S., Lachieze-Rey M., Henriksen R.N., 1993,
Astron. Astrophys., {\bf{273}}, 357}
\item{Riess A.G., Press W.H., Kirschner R.P., 1995a, Ap.J. Lett.
{\bf{438}}, L17}
\item{Riess A.G., Press W.H., Kirschner R.P., 1995a, Ap.J. Lett.
{\bf{445}}, L91}
\item{Rubin V.C., Thonnard N., Ford W.K., Roberts M.S., 1976,
Astron. J., {\bf{81}}, 719}
%
\item{Saha A., Labhart L., Schwengeler H., Machetto F.D., Panagia N.,
Sandage A., Tammann G.A., 1994, Ap.J., {\bf{425}}, 14}
\item{Saha A., Sandage A., Labhart L.,Schwengeler H., Tammann G.A.,
Panagia N., Machetto F.D., 1995, Ap.J., {\bf{438}}, 8}
\item{Salucci P., Frenk C.S., Persic M., 1993, MNRAS, {\bf{262}}, 392}
\item{Sandage A., 1994, Ap.J., {\bf{430}}, 13}
\item{Sandage A., Tammann G., 1975a, Ap.J., {\bf{196}}, 313}
\item{Sandage A., Tammann G., 1975b, Ap.J., {\bf{197}}, 265}
\item{Sandage A., Tammann G., 1993, Ap.J., {\bf{415}}, 1}
\item{Schechter P.L., 1980, Astron.J., {\bf{85}}, 801}
\item{Schmidt B.P., Kirschner R.P. Eastman R.G., 1992, Ap.J., {\bf{395}},
366}
\item{Schmidt B.P., et al, 1994, Ap.J., {\bf{432}}, 42}
\item{Simmons J.F.L., Newsam A.M., Hendry M.A., 1995, Astron. Astrophys.,
{\bf{293}}, 13}
\item{Shaya E.J., Tully R.B., Pierce M.J., 1992, Ap.J., {\bf{391}}, 16}
\item{Shaya E.J., Tully R.B., Peebles P.J.E., 1995, Ap.J., in press}
\item{Sodre Jr.L., Lahav O., 1993, MNRAS, {\bf{260}}, 285}
%
\item{Teerikorpi P., 1993, Astron. Astrophys., {\bf{280}}, 443}
\item{Tonry J.L., Schneider D.P., 1988, A.J., {\bf{96}},807}
\item{Triay R., Lachieze-Rey M., Rauzy S., 1994, Astron. Astrophys.,
{\bf{289}}, 19}
\item{Tully R.B., Fisher J.R., 1977, Astron. Astrophys.,
{\bf{54}}, 661}
%
%
\item{van den Bergh, S., 1995, JRASC, {\bf{89}}, 6}
\item{Vaughan T.E., Branch D., Miller D.L., Perlmutter S.,
1995, Ap.J., {\bf{439}}, 558}
\item{Visvanathan N., Sandage A., 1977, Ap.J., {\bf{216}}, 214}
%
\item{Willick J., 1994, Ap.J. Supp., {\bf{92}}, 1}
%
%
%
\item{Zaccheo T.S., Gonsalves R.A., Ebstein S.M., Nisenson P.,
1995, Ap.J. Lett., {\bf{439}}, L43}
\end{itemize}
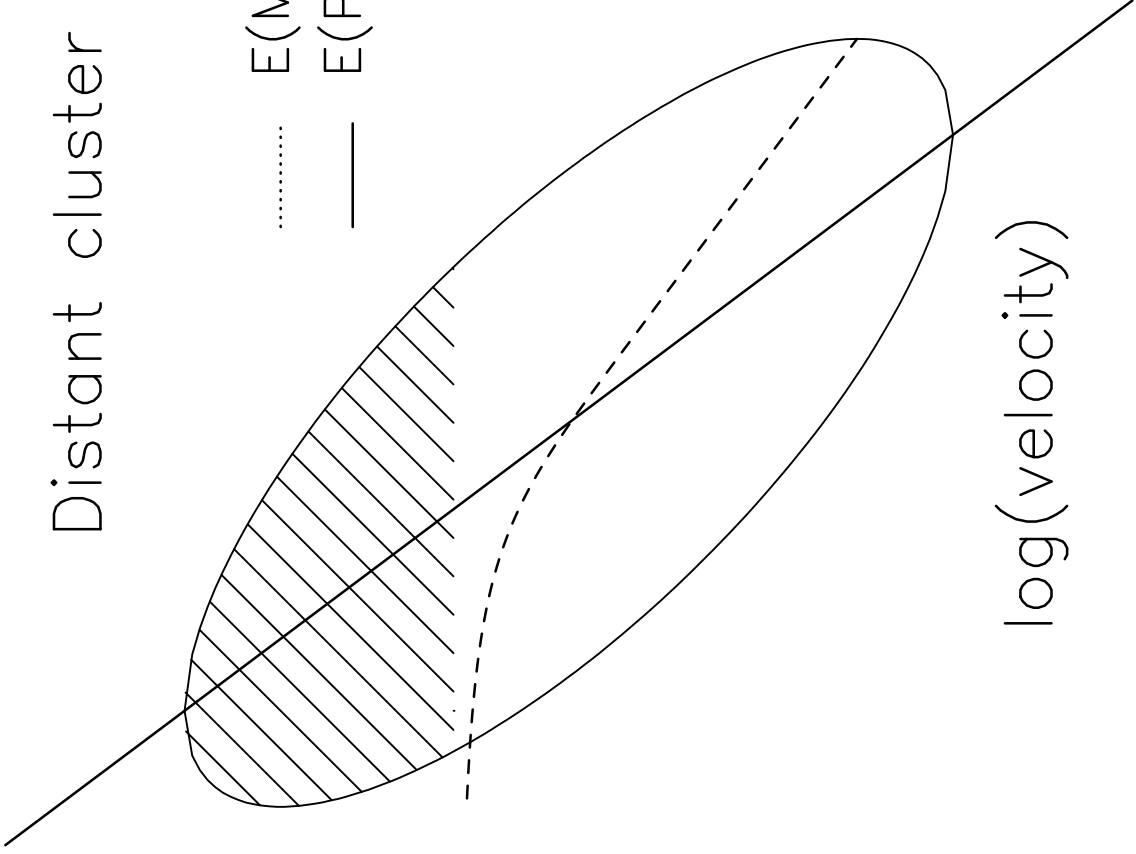
\end{document}